\documentclass{nature}
\usepackage{epsfig}
\usepackage{amssymb}

\newcommand{\Msolar}{M$_{\odot}$}

\newcommand{\degr}{$^{\circ}$}

\title{A mass transfer origin for blue stragglers in NGC 188 as revealed by half-solar-mass companions}

\author{Aaron M. Geller $^{1,2}$~~\&  Robert D. Mathieu$^{1}$,}

\begin{document}

\maketitle

\begin{affiliations}
 \item Department of Astronomy, University of Wisconsin - Madison, WI 53706 USA
 \item Center for Interdisciplinary Exploration and Research in Astrophysics (CIERA) \& Department of Physics and Astronomy, Northwestern University, 2145 Sheridan Rd, Evanston, IL 60208, USA. 
\end{affiliations}

\begin{abstract}
In open star clusters, where all members formed at about the same time, blue straggler stars are typically observed to be brighter and 
bluer than hydrogen-burning main-sequence stars, and therefore should already have evolved into giant stars and stellar remnants.
Correlations between blue straggler frequency and cluster binary fraction\cite{sol08}, core mass\cite{kni09}, and radial 
position\cite{fer97} suggest  
that mass transfer or mergers in binary stars dominates the production of blue stragglers in open clusters.
Analytic models\cite{leo96,lei11}, detailed observations\cite{san05}, 
and sophisticated $N$-body simulations\cite{hur05}, however, argue in favor of stellar collisions.
Here we report that the blue stragglers in long-period binaries in the old\cite{sar99} (7 Gyr) open cluster NGC 188 have companions 
with masses of about half a solar mass, with a surprisingly narrow mass distribution.
This conclusively rules out a collisional origin, as the collision hypothesis predicts a companion-mass distribution with significantly higher masses.
Mergers in hierarchical triple stars\cite{per09} are marginally permitted by the data, but the observations do not favor this hypothesis.  
The data are closely consistent with a mass transfer origin for the long-period blue straggler binaries in NGC 188, in which the 
companions would be white dwarfs of about half a solar mass.
\vspace{1em}

\end{abstract}



The NGC 188 blue stragglers have a very high binary frequency of $76 \pm 19$\% (for periods $< 10^4$ days).\cite{mat09}
The orbital period distribution is remarkable; 12 of the 16 blue straggler binaries have periods of order 1000 days, 
and all but two of the blue straggler binaries have periods longer than 100 days.  The two short-period blue 
straggler binaries show evidence for binary encounters were involved in their formation.\cite{mat09}
We focus here on the "long-period" blue straggler binaries, whose orbital solutions yield periods longer than 100 days.

\begin{figure}

\begin{center}
\includegraphics[width=0.9\linewidth]{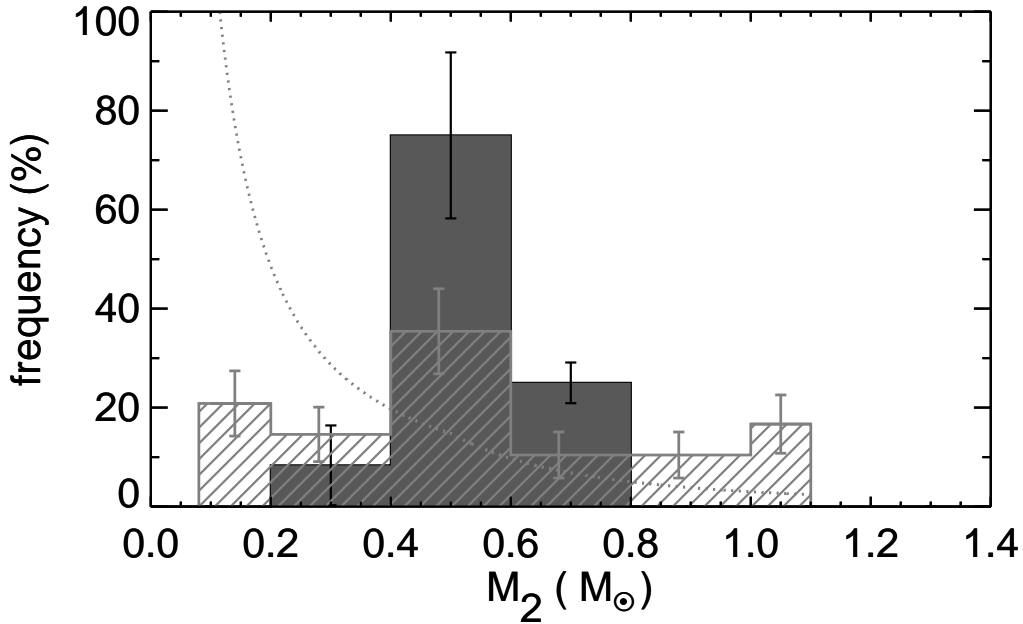}
\end{center}


\caption{\textbf{Companion-mass distribution for the 12 blue straggler binaries in NGC 188 with periods of order 1000 days.}
The distribution is peaked near the typical Carbon-Oxygen white dwarf 
mass of 0.55~\Msolar,  consistent with theoretical predictions from the Case C mass-transfer hypothesis.
We use a statistical method\cite{maz92} to derive the companion-mass distribution from the observed mass functions
(see Supplementary Information).  
To do so, we first estimate masses for the blue stragglers based on standard
stellar evolutionary tracks\cite{mar08}, and assume an isotropic inclination distribution.
The shaded histogram shows the resulting companion-mass distribution and is normalized to show
the frequency.
The error bars on the histogram show the 95\% confidence intervals and are converted from the Poisson uncertainties on the 
mass-function distribution using a Monte Carlo analysis.
We note that standard evolutionary tracks may underestimate\cite{mat09} the mass of blue stragglers by up to about 15\%.
Accounting for this potential bias does not change the results found here (nor in Figure~\ref{fmfig}).
For comparison we also plot a standard initial mass function for single stars\cite{kro01a} (dotted gray line)
for companion masses between 
0.08~\Msolar~and 1.1~\Msolar~(from the Hydrogen burning limit to the current main-sequence turn-off mass in NGC 188).
The gray hatched histogram shows the observed tertiary-mass distribution\cite{tok97} evolved to 7 Gyr in isolation\cite{hur00}.  
The lowest-mass bin extends from 0.08~\Msolar~to 0.2~\Msolar, 
and the highest-mass bin extends from 1.0~\Msolar~to 1.1~\Msolar.  Both of these bins are renormalized to reflect the different bin sizes.

\label{obsfig}
}
\end{figure}

In Figure~\ref{obsfig} we show the companion-mass distribution for the twelve NGC 188 blue straggler binaries with periods of order 1000 days.
The orbital solutions are derived from spectroscopic data\cite{mat00,gel09} obtained in the WIYN
Open Cluster Study (WOCS).  Because we do not detect the flux from the companions to these blue stragglers, 
the orbital solutions provide mass functions rather than mass ratios.
We therefore use a statistical algorithm\cite{maz92}
to convert the mass functions to the companion-mass distribution shown here (see Supplementary Information). 
The distribution is narrow and peaked with a mean of 0.53~\Msolar~and a mode of 0.5~\Msolar.

\begin{figure}

\begin{center}
\includegraphics[width=0.69\linewidth]{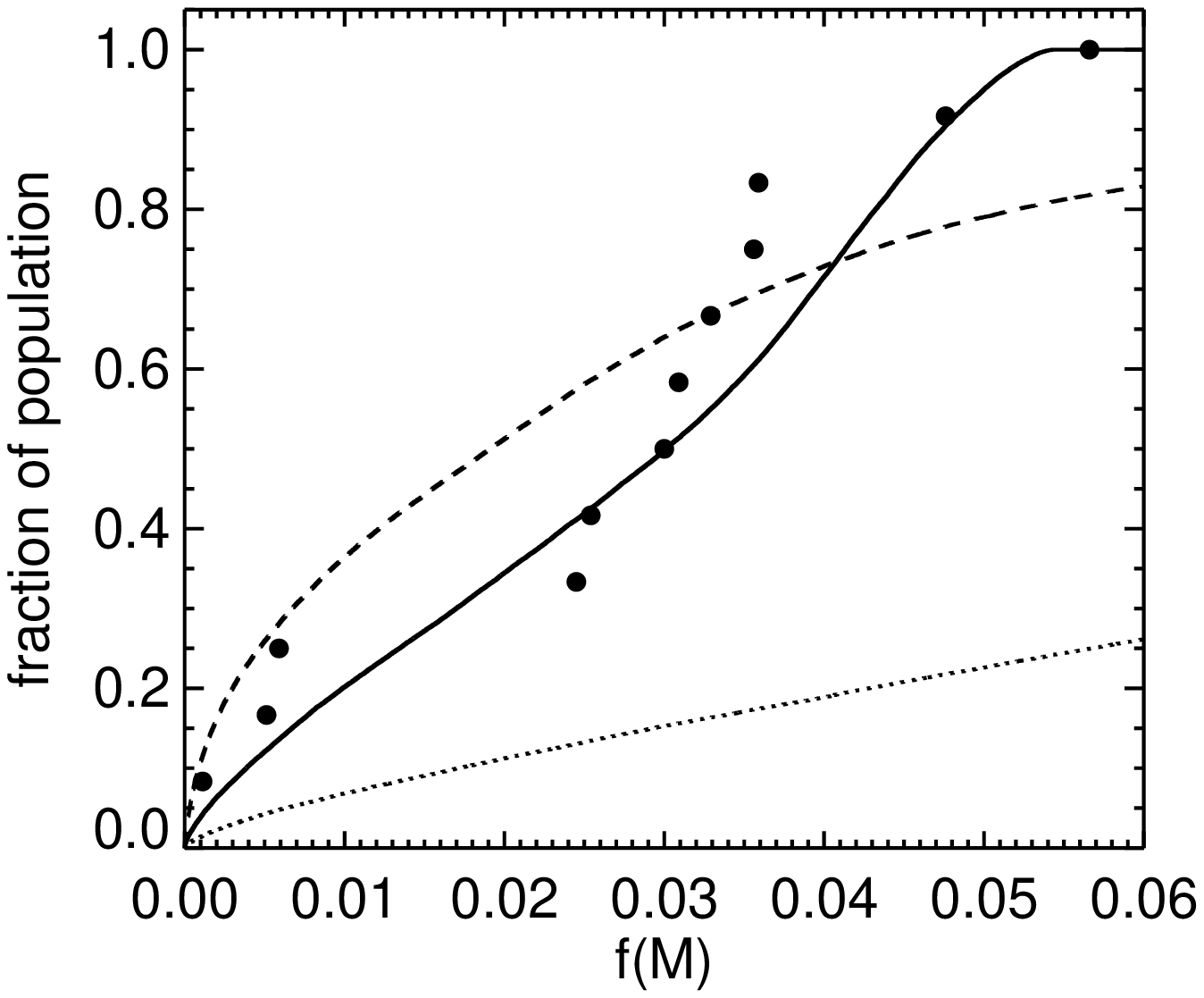}
\end{center}

\caption{\textbf{Cumulative distribution of mass functions of the NGC 188 blue straggler binaries with periods of order 1000 days.}
The black points show the observed mass functions derived directly from the kinematic orbital solutions\cite{gel09}, defined as:
\begin{equation}
f(M_1,M_2,i) = \frac{M_2^3}{(M_1+M_2)^2} \sin^3 i
\end{equation}
where $M_1$ is the mass of the primary star (here the blue straggler), 
$M_2$ is the mass of the companion star and $i$ is the inclination of the orbit to our line of sight. 
(Orbits that are edge-on have inclinations of 90\degr.)
To test the blue straggler formation hypotheses, we compare this observed distribution to three theoretical mass-function distributions,
all derived using our blue straggler mass estimates and assuming isotropically distributed inclination angles.
The solid line shows the resulting distribution assuming all companions have masses of 0.55~\Msolar, the typical Carbon-Oxygen white 
dwarf mass predicted by the Case C mass-transfer hypothesis. 
The dotted line shows the mass-function distribution of the predicted companion masses for blue straggler binaries 
formed by collisions in the NGC 188 $N$-body model. 
Finally the dashed line shows the distribution derived by drawing companion masses from the evolved tertiary-mass distribution 
shown in Figure~\ref{obsfig}.
A K-S test rules out the collisional hypothesis at the $>$99\% confidence level.  
A K-S test does not rule out the merger hypothesis. However there is only a 6.6\% chance that all companions 
would be undetected in our spectra if drawn from the evolved tertiary-mass distribution, and only a 1.8\% chance that these 
companions would also realize the observed mass-function distribution shown here.
The observed mass-function distribution is statistically indistinguishable from that predicted by the mass-transfer 
hypothesis, and all white dwarf companions would be undetectable in our spectra, given their low luminosities.
\label{fmfig}
}
\end{figure}

Predictions for the companions to blue stragglers resulting from mass transfer in solar-type stars are well established by theory.  
Case C mass transfer (from an asymptotic-giant to a main-sequence star) leaves a Carbon-Oxygen white dwarf companion in an orbit of order 
1000 days with a mass of about 0.5~\Msolar~to 0.6~\Msolar, 
dictated by the core mass of the asymptotic-giant donor at the end of the mass-transfer phase.\cite{mcc64,hur02,bel08,che08} 
This prediction is qualitatively reproduced by the NGC 188 blue straggler companion-mass distribution shown in Figure~\ref{obsfig}.
To check quantitatively for consistency we compare the observed mass-function distribution (points in Figure~\ref{fmfig}) 
to a theoretical mass-function distribution derived assuming all companions have the typical Carbon-Oxygen white dwarf mass,
0.55 \Msolar~(solid line in Figure~\ref{fmfig}).
A Kolmogorov-Smirnov (K-S) test shows that the theoretical and observed distributions are indistinguishable. 

Additionally NGC 188 contains one blue straggler in a binary with an orbital period of about 120 days, 
the companion flux of which we also do not detect in our spectra.
The observed mass function and orbital period for this system are consistent with the theoretical predictions of Case B mass transfer
(from a giant to a main-sequence star), 
where a Helium white dwarf companion with a mass of about 0.25~\Msolar~to 0.5~\Msolar~is expected.\cite{mcc64,hur02,bel08,che08}
Thus the companions to all long-period blue straggler binaries in NGC 188 are consistent with a mass-transfer origin.

\begin{figure}

\begin{tabular}{lll}
\hspace{-4em} \includegraphics[width=0.32\linewidth]{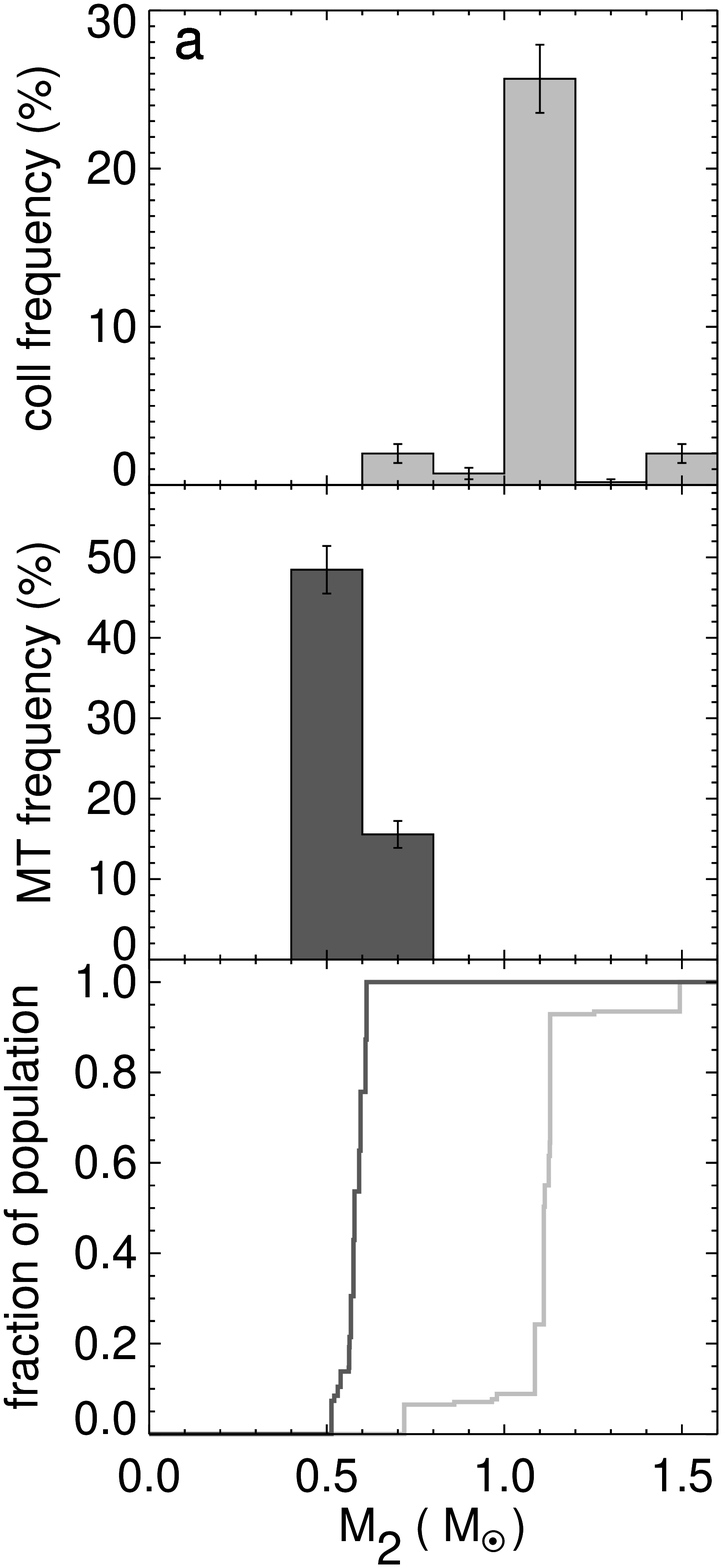} & \includegraphics[width=0.32\linewidth]{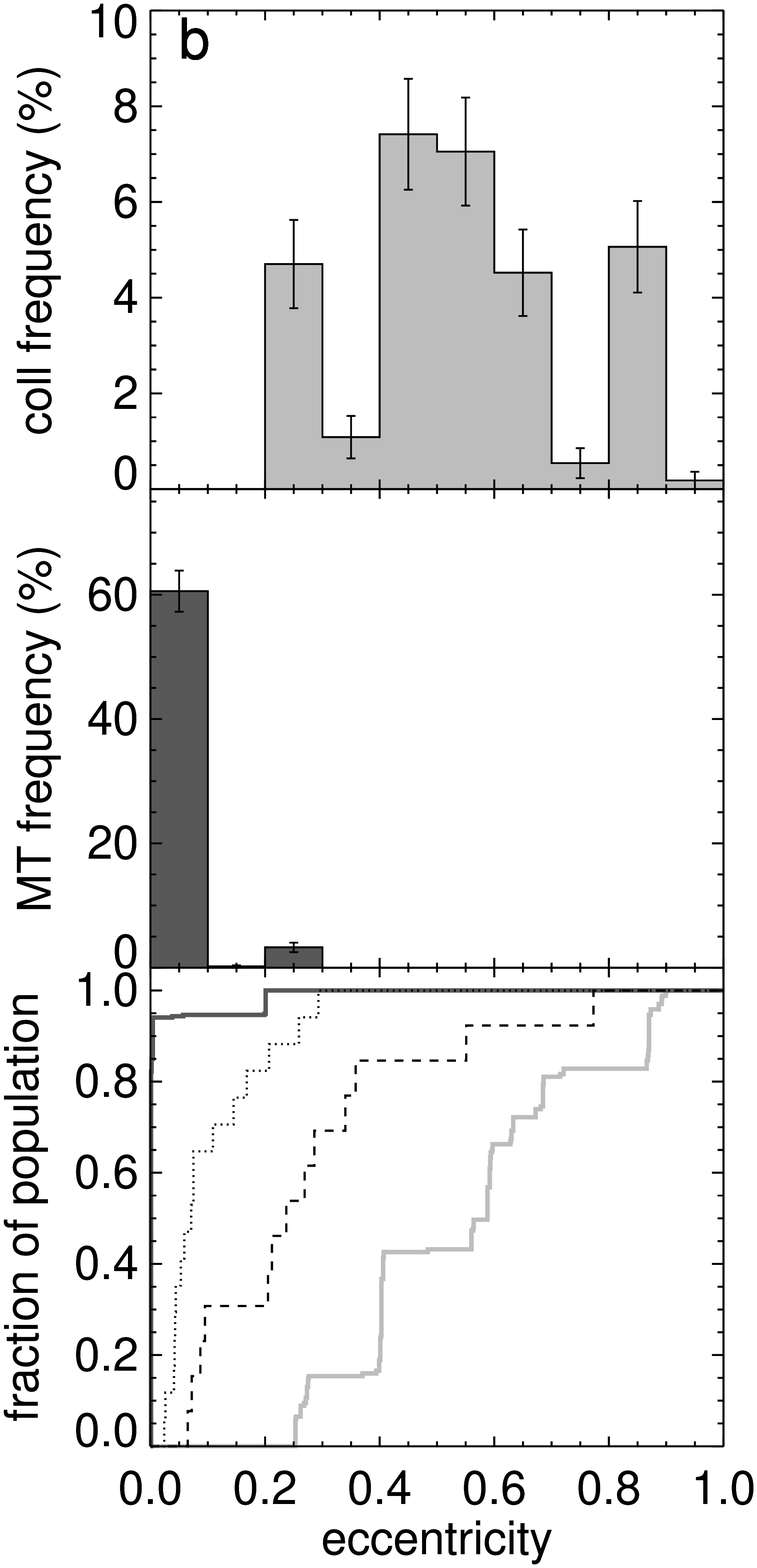} & \includegraphics[width=0.32\linewidth]{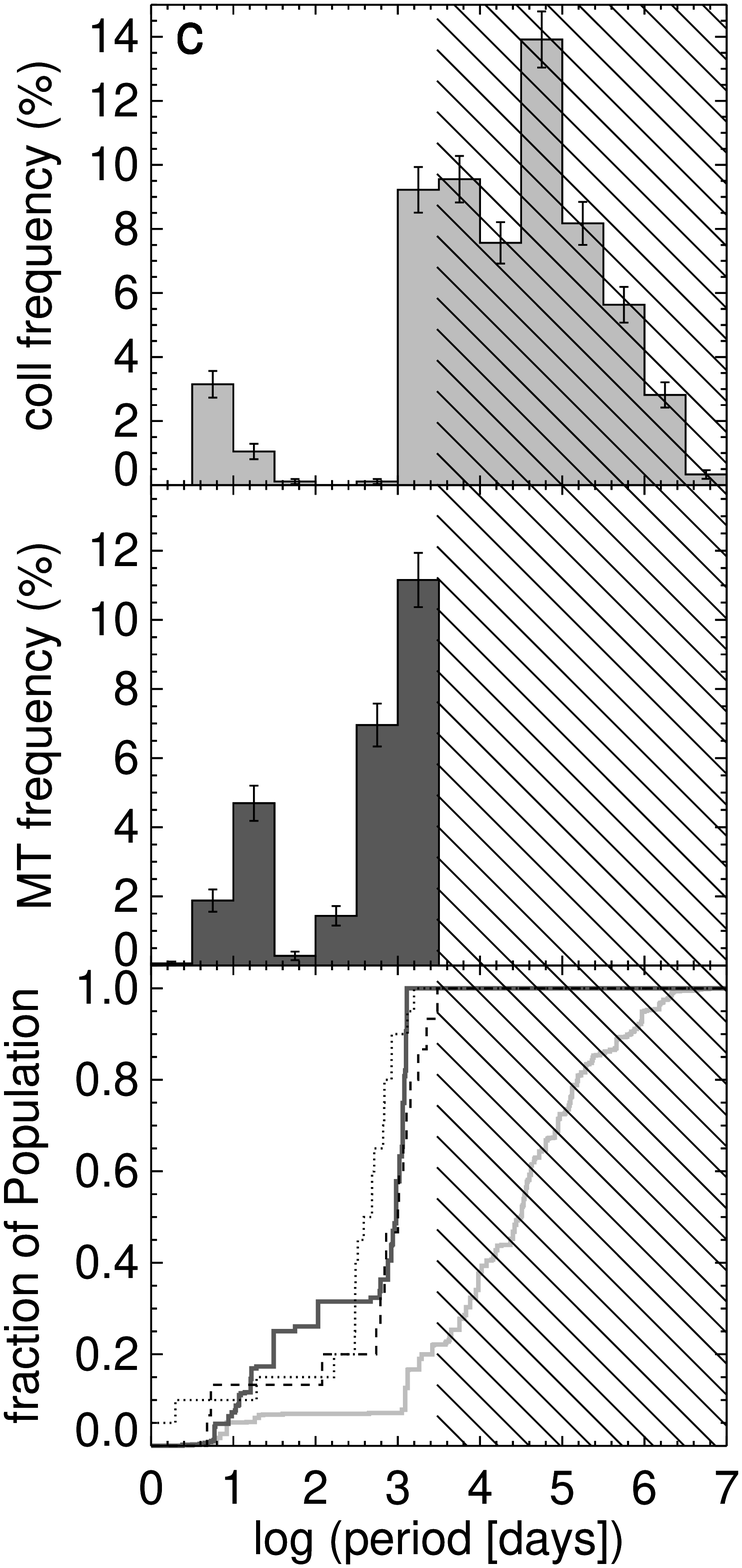} \\
\end{tabular}

\caption{\textbf{Distributions of binary orbital elements for the blue stragglers in the NGC 188 $N$-body model.}
\textbf{a}, Companions mass; \textbf{b}, orbital eccentricity; \textbf{c}, orbital period.
The top plots show blue stragglers formed in collisions (coll), and the middle plots show blue stragglers created through mass transfer (MT).
The bottom plots compare the cumulative distributions for these two populations, with collision products in light gray 
and mass-transfer products in dark gray. 
The sample contains all blue straggler binaries present from 6 - 7.5 Gyr in the model, spanning the
uncertainty in the cluster age.  
In \textbf{a} and \textbf{b}, we show only blue straggler binaries with periods between 100 and 3000 days (the period range of our NGC 188 long-period 
blue straggler binaries). In \textbf{c}, we include blue straggler binaries of all periods (with the hatched regions indicating binaries at periods 
beyond our detection limit).  
Each bin is normalized by the total number of blue straggler binaries in the sample, and the error bars show the Poisson counting uncertainties. 
Additionally, in the bottom panels of \textbf{b} and \textbf{c} we show the cumulative eccentricity (\textbf{b}) and period (\textbf{c}) 
distributions of the blue straggler 
binaries in NGC 188 (dashed lines) and the field\cite{car05} (dotted lines).
The $N$-body model predicts that collision products will have significantly more massive companions, larger eccentricities and 
longer periods than the blue stragglers in NGC 188. 
However, the predicted companion masses and periods for mass-transfer blue stragglers are consistent with those of the 
long-period NGC 188 blue straggler binaries.
We note that the $N$-body model does not implement proposed ``eccentricity-pumping'' mechanisms that are required to reproduce 
the non-zero eccentricities observed for post-mass-transfer systems, including the field blue stragglers. 
Thus the predicted eccentricities for the mass-transfer products are uncertain, although still likely to be lower than those of the collision products.
\label{simfig}
}
\end{figure}

We investigate the predicted companion-mass distribution from the collision hypothesis using a sophisticated $N$-body model of NGC 188 that 
incorporates detailed stellar and binary evolution with stellar dynamics, and thereby produces blue stragglers through both collisions and 
mass-transfer processes (see Supplementary Information).
The resulting distributions of companion mass, eccentricity and period for the blue stragglers in binaries that respectively formed 
through the collision and mass-transfer mechanisms (Figure~\ref{simfig}) show marked differences.
We focus here on the simulated blue stragglers in binaries with periods of $100 - 3000$ days (matching the period range of our NGC 188 long-period 
blue straggler binaries).

The very narrow distribution and mean mass of 0.58~$\pm$~0.01~\Msolar~for companions to the mass transfer blue stragglers in the $N$-body model
are in good agreement with the population synthesis predictions discussed above.
However, the mean mass of companions to collisionally formed blue stragglers is 
1.11~$\pm$~0.02~\Msolar, nearly twice that of the mass-transfer blue stragglers.
This result 
demonstrates the finding that dynamical exchanges are more likely to insert a higher mass member of the encounter into the binary.\cite{hur02b}

These differences in companion mass 
reflect profound differences in the evolutionary states of the companions resulting from each process. 
All of the simulated blue straggler binaries that formed through mass-transfer processes have white dwarf companions, 
the remnants of the giant star donors.  Fewer than 1\% of the blue stragglers in binaries that formed through collisions have white-dwarf
companions, whereas 80\% have main-sequence companions (and the remaining have giant or blue straggler companions).

A K-S test rules out at the $>$99\% confidence level the hypothesis that
the observed mass functions for the long-period NGC 188 blue straggler binaries 
are drawn from a parent population of collisional origin (dotted line in Figure~\ref{fmfig}).
Additionally collision products are predicted to have significantly higher eccentricities and longer periods
than are observed for the NGC 188 blue stragglers (both at the $>$99\% confidence levels, respectively).
We therefore rule out the hypothesis that the long-period NGC 188 blue straggler binaries have an origin in collisions.

Three of the long-period NGC 188 blue stragglers have measured
orbital eccentricities consistent with circular orbits. By contrast,
no NGC 188 solar-type main-sequence binaries in this period range
have circular orbits. These blue straggler circular orbits are
suggestive of a mass-transfer origin. Rapid tidal circularization of the
orbit during mass transfer has been a long-held expectation, as is
seen in the predicted eccentricity distribution for mass-transfer
products in the NGC 188 model (Figure~\ref{simfig}b).

However observational and theoretical evidence suggests that mass transfer will not always lead to circular orbits.
Proposed ``eccentricity-pumping mechanisms'' address this issue and are under development.\cite{sok00,bon08,sep09}  
Thus the theoretical eccentricity distribution for blue stragglers formed by mass transfer is uncertain. 

Fortuitously, the blue stragglers of the Galactic field provide a basis for empirical comparison\cite{car05}. 
Specifically we compare the NGC 188 blue stragglers with a blue straggler sample identified within the population of metal-poor 
thick-disk and halo stars, which is found to be coeval.\cite{car89}
These field blue stragglers probably formed in 
isolation, presumably through mass transfer processes\cite{car05}. In fact, the field blue stragglers are observed to have a high binary
frequency, a period distribution that peaks near a few 100 - 1000 days (dotted line in Figure~\ref{simfig}c), and a mean companion mass consistent with 
0.55~\Msolar, all again consistent with a mass transfer origin.
 
These long-period field blue straggler binaries show a range of non-circular eccentricities (dotted line in Figure~\ref{simfig}b).
Enhancement of eccentricity through subsequent dynamical encounters
cannot explain the non-circular orbits for field blue stragglers, due to the low stellar density of the Galactic field.
Therefore, if the long-period binaries among the field blue stragglers were formed through mass transfer, this is further evidence for the 
existence of an eccentricity-pumping mechanism.

The eccentricity distribution of the long-period NGC 188 blue straggler binaries is shifted to higher eccentricities than that of the long-period 
field blue straggler binaries.  
However, owing to the higher densities in the cluster core, dynamical encounters may have increased the eccentricities of the NGC 188 blue stragglers.
If we exclude from the analysis blue stragglers in NGC 188 within 1.5 core radii from the cluster center, the two eccentricity distributions are 
statistically indistinguishable. 
The similarity in periods, companion masses and eccentricities would be a natural consequence of both the field and NGC 188 long-period blue straggler 
binaries being formed by mass-transfer processes.

Finally, we investigate the possibility of blue straggler formation through mergers in hierarchical triples\cite{iva08,per09}.
The potential progenitors of the blue stragglers produced through this mechanism would be triples with short-period ($\lesssim10$~days) 
inner binaries having a total mass between 1.2~\Msolar~and 2.2 \Msolar. (2.2 \Msolar~is twice the turnoff mass in NGC 188.)
The frequency of dynamically formed triples in the NGC 188 model with these orbital parameters is never 
high enough to contribute considerably to blue straggler production through this mechanism.
If the merger mechanism is important, triples must form primordially with suitable orbital parameters in cluster environments.

Observationally, the triple population in open clusters is poorly known, but triples are common in the field.\cite{tok06}
Field triples in the Multiple Star Catalogue\cite{tok97} 
with measured masses indicating inner binaries of total mass between 1.2 and 2.2 \Msolar~and inner orbital periods less than 10 days
have a nearly uniform tertiary-mass distribution populated by main-sequence stars.
However, the typical age of local field stars is 4 Gyr.\cite{rob03}  
If we evolve the tertiary-mass distribution to 7 Gyr in isolation\cite{hur00}, 
15\% of the tertiaries evolve to become white dwarfs.  The gray hatched histogram in Figure~\ref{obsfig} shows the resulting
tertiary-mass distribution at 7 Gyr.
The mass distribution is qualitatively broader than that of the companions to the NGC 188 long-period blue stragglers.
However a K-S test comparing the mass-function distributions does not rule out the triple hypothesis.

The fact that we do not detect in our spectra the flux from companions to any of the long-period blue straggler binaries also
constrains the companion masses. 
The higher-mass main-sequence stars in the evolved tertiary-mass distribution (Figure~\ref{obsfig}) would be easily detected
if these were the true companions to the long-period blue straggler binaries in NBC 188.
A Monte Carlo analysis yields a 6.6\% probability that all companions to the long-period NGC 188 blue straggler binaries
would be undetected in our spectra if drawn from the evolved tertiary-mass distribution, and only a 1.8\% probability that 
these companions would also realize the observed mass-function distribution of the long-period NGC 188 blue straggler binaries
(see Supplementary Information).
Thus mergers in hierarchical triples are not favored by the observations, but the data are not sufficient to rule out this hypothesis completely.


We aim to detect directly the flux from the white dwarf companions predicted by the mass transfer mechanism with forthcoming Hubble
Space Telescope observations of the NGC 188 long-period blue straggler binaries in the ultraviolet. 
These observations will be invaluable for distinguishing between the two remaining formation hypotheses: binary mass transfer and mergers 
in hierarchical triples.

\vspace{3em}


\bibliography{GellerMathieu2011}

\begin{addendum}
\item[Supplementary Information] is linked to the online version of the paper at www.nature.com/nature.
 \item We thank the staff of the WIYN Observatory and the many graduate and undergraduate students who have assisted in observing NGC 188. 
Thanks to J.~Hurley, A.~Sills, N.~Leigh, R.~Taam and H.~Perets for their comments and suggestions.
Both A.M.G and R.D.M. were visiting astronomers at Kitt Peak National Observatory, National Optical Astronomy Observatory, which is operated by 
the Association of Universities for Research in Astronomy (AURA) under cooperative agreement with the National Science Foundation. 
The WIYN Observatory is a joint facility of the University of Wisconsin –- Madison, Indiana University, Yale University and the US National Optical 
Astronomy Observatories.  
This work was funded by the US National Science Foundation grant AST-0908082 to the University of Wisconsin - Madison, 
the Wisconsin Space Grant Consortium and the Lindheimer Fellowship at Northwestern University.
 \item[Author Contributions] A.M.G. and R.D.M. contributed equally to this work.
 \item[Author Information] Reprints and permissions information is available at www.nature.com/reprints.  The authors have no competing financial interests. Correspondence and requests for materials should be addressed to A.M.G. (a-geller@northwestern.edu).

\end{addendum}

\pagebreak


\pagebreak

\pagebreak

\pagebreak
\textbf{Supplementary Information}


\section{Deriving the Companion-Mass Distribution}

We do not detect the flux from the companions to any of the long-period blue straggler binaries in NGC 188.
Thus our kinematic orbital solutions return mass functions.
In order to convert the observed distribution of mass functions into a distribution of companion masses we 
use an iterative statistical algorithm\cite{maz92}.  

First we require estimates for the primary masses of the blue stragglers, which we derive by comparing the 
observed luminosity and temperature of each blue straggler to theoretical stellar evolutionary tracks\cite{mar08}.
This analysis shows that the long-period blue straggler binaries in NGC 188 have masses between 1.2~\Msolar~and 1.6~\Msolar.
We note that standard evolutionary tracks may underestimate the mass of blue stragglers by up to about 15\%,
as found for the two short-period blue stragglers in NGC 188.\cite{mat09}
However the differences are small enough to neglect for the purposes of this statistical distribution.

We begin the algorithm with an initial guess of a uniform companion-mass distribution. 
Then for each binary we assign a distribution of inclination angles over the range allowed by 
each mass function and primary-mass estimate, respectively.  The inclinations are first chosen to be isotropically distributed and 
then multiplied by a correction factor that is a function of the assumed companion-mass distribution.
Using the mass functions, the primary-mass estimates and the derived inclination distributions we create a sample of 
synthetic binaries for each observed binary.  By summing over all synthetic binary samples we derive the next estimate 
for our companion-mass distribution.  This distribution is then used as input for the next iteration of the algorithm, and 
the process is repeated until the solution converges.

This method does not determine the individual companion masses for any given binary, but instead derives the 
distribution of companion masses for a given population of binaries.  The result for the blue straggler binaries in NGC 188 
with orbital periods of order 1000 days is the peaked distribution shown in Figure~\ref{obsfig}, with a mean of 0.53~\Msolar.  

For Figure~\ref{obsfig} (and Figure~\ref{fmfig}) we are only interested in the blue straggler binaries in NGC 188 with periods of order 1000 days.
Thus we exclude the one blue straggler binary with an orbital period of about 120 days, the one blue straggler binary without an 
orbital solution and the two short-period blue stragglers for which we detect the companions in our spectra.

This distribution is not sensitive to the blue straggler mass used in the analysis.
If we simply assign the same mass to each blue straggler within the range of 
1.2~\Msolar~to 1.6~\Msolar~derived above, the resulting distributions stay well within the uncertainties shown in the figure.
If we take the extreme case of setting all blue stragglers to twice the turnoff mass (2.2 \Msolar),  
the general form of the distribution 
remains the same, but the peak shifts towards more massive companions by about 0.2~\Msolar, roughly equal to the 
standard deviation of a Gaussian fit to the distribution shown in Figure~\ref{obsfig}.  
However no current model predicts a population of such massive blue straggler.

Finally, we note that the statistical algorithm is unable to fully resolve sharp features (e.g., delta functions), a property
regarded as the intrinsic ``instrumental profile'' of the technique\cite{maz92}, which may partly explain the tails of the derived distribution.

\section{NGC 188 \textit{N}-body Model}

The blue stragglers in the NGC 188 $N$-body model are derived from twenty nearly identical simulations, each 
starting with the same parameters (e.g., total cluster mass, binary frequency, distribution of binary orbital parameters, etc.)
but with different randomized stellar positions, velocities, masses, binary orbital elements, etc.
Each simulation uses the \texttt{NBODY6} code\cite{aar03} to model the dynamical evolution, with 
stellar\cite{hur00} and binary\cite{hur02} evolution included.
We have made slight modifications to this code to define the initial binary population and output format.
Importantly the model contains detailed binary evolution prescriptions for tidal circularization 
and synchronization, angular momentum loss mechanisms (e.g., magnetic braking and gravitational radiation), mass transfer from
Roche lobe overflow, accretion from stellar winds, common-envelope events, mergers and
direct stellar collisions, thus providing numerous pathways to create blue stragglers.\cite{hur05}

The modeling of blue stragglers is given special attention.\cite{hur02,hur05}
In short, if a main-sequence star gains mass (e.g., through mass-transfer processes or collisions), it evolves up
along the main sequence to higher luminosity and effective temperature.  The lifetime of the
blue straggler prior to becoming a giant is determined based on the fraction of unburned hydrogen 
remaining in the core of the star, which can be replenished when the star gains mass as determined by the details of 
the formation mechanism. Within the simulation we identify blue stragglers as being more massive
than a normal main-sequence star at the turnoff, while still maintaining the structure and evolutionary
state of a main-sequence star.  The combination of twenty simulations helps reduce the effects of the stochastic nature
of blue straggler formation in $N$-body simulations.

The initial parameters of any $N$-body open cluster simulation, and particularly those of the binary population,
have a great impact on the dynamical evolution of the cluster and the production rates
and mechanisms for blue stragglers (and other anomalous stars).  
Where possible we have attempted to base all initial parameters directly on observations.
Importantly, we use detailed observations of the binary population in a young open cluster (M35; 150 Myr)
to define the initial binary frequency and distributions of orbital parameters.\cite{gel10b,gel10}
After 7 Gyr of evolution, the model matches the observed mass, core and tidal radii, and the main-sequence 
binary frequency and distributions of binary orbital parameters in NGC 188 in detail.

To investigate the blue stragglers (Figure~\ref{simfig}) we first identify the formation mechanisms for each 
individual blue straggler by examining their dynamical histories, respectively. 
Collisions in the NGC 188 model generally occur at periastron within a binary driven to high eccentricity by a dynamical encounter.
Mass transfer occurs through either the Case C or Case B mechanisms as discussed in the main text. 
(Low-mass blue stragglers also may form in long-period binaries if a main-sequence star accretes sufficient mass 
from the stellar wind of a giant star companion.\cite{hur02} However this mechanism cannot produce blue stragglers 
as luminous and presumably massive as the long-period blue straggler binaries in NGC 188.)

For the analysis in this paper, we include all blue stragglers present at each time step in each of the twenty simulations between
the cluster age of 6 - 7.5 Gyr, covering the range in ages derived for NGC 188 in the literature. 
In so doing, we effectively weight the results by the lifetime of each blue straggler in a given orbital configuration.

The NGC 188 model assumes that tidal dissipation during mass transfer will rapidly
circularize the binary orbit.  Therefore nearly all blue straggler binaries in the NGC 188 model that formed through mass transfer have orbital
eccentricities of zero. The very few non-circular orbits are the result of dynamical encounters after the formation of the blue stragglers. 
However observations of post-mass-transfer systems show a wide 
range of non-zero eccentricities (e.g. barium-star systems\cite{jor98} and post-asymptotic-giant binaries\cite{van99}).
These new theories have yet to be included in $N$-body models.  Thus the eccentricities for the mass transfer 
blue stragglers in the NGC 188 model are uncertain, although still likely to be lower than those of the collision products.

\section{Monte Carlo Analysis}

As we do not detect the flux of the companions to any of the long-period NGC 188 blue stragglers in our spectra,
we cannot directly determine their evolutionary states.
If the dominant formation channel is mergers in hierarchical triples, then the companions to the blue stragglers would 
be the original tertiary stars, and would be drawn from the observed tertiary-mass distribution\cite{tok97} that 
has been evolved in isolation for 7 Gyr (shown in Figure~\ref{obsfig}).
This evolved tertiary-mass distribution contains triples that would be the likely progenitors of the long-period blue straggler binaries
in NGC 188, having inner binaries of a total mass consistent with that of the NGC 188 blue stragglers and orbital periods of $<$10 days.

A K-S test comparing the observed mass-function distribution for the long-period blue straggler binaries in NGC 188 to that of the 
evolved tertiaries yields a distance statistic of 0.27 (which corresponds to a probability of 36\%).
Here we also determine the likelihood that such companions would produce thirteen binaries without any detected light 
from the companions.

To test this hypothesis we perform a Monte Carlo analysis in which we assume all companions are main-sequence stars
with masses between 0.08~\Msolar~and 1.1~\Msolar~(from the Hydrogen burning limit to the current main-sequence turnoff mass in NGC 188).
Given WIYN spectra of an NGC 188 binary, we have the ability to detect the flux from any companion that is at most 
10 times less luminous than the primary.\cite{gel09}  This sets our detectability limit for the simulated companions. 

We create 10$^6$ realizations of the NGC 188 long-period blue straggler binary population.
For each realization we choose random companions for these thirteen NGC 188 blue stragglers from the evolved tertiary-mass distribution.
We note that 15\% of the tertiaries in this distribution are expected to be Carbon - Oxygen white dwarfs with masses of about 0.55~\Msolar.
Both white dwarf and main-sequence star companions of this mass would be undetected in our spectra.

We then check the luminosity differences between each blue straggler and its simulated companion using a 7 Gyr Padova isochrone\cite{mar08}.  
None of the long-period NGC 188 blue stragglers have detectable companions.  Therefore we reproduce this result if zero of the 
thirteen simulated blue straggler binaries in a given realization are detectable.  
Only 6.6\% of the realizations of the Monte Carlo blue straggler binary population satisfy this criterion, which we will call $P(A)$.

Finally, we wish to calculate the probability that all blue stragglers will be undetected in our spectra ($P(A)$) and the companions 
will reproduce the observed mass-function distribution of the long-period blue straggler binaries in NGC 188 ($P(B)$). 
These two probabilities are dependent and therefore $P(A$~and~$B) = P(A) \times P(B|A)$. 
If the K-S distance statistic found when comparing the mass-function distribution of the evolved tertiaries
to that of a given realization
is greater than or equal to that found above for the  NGC 188 blue stragglers (i.e., 0.27), the realization reproduces the 
observations.  Of the 6.6\% of realization that have zero detectable companions, 26.9\% also reproduce the observed mass-function 
distribution ($P(B|A)$).  Therefore we find a 1.8\% probability that we detect zero companions \textit{and} realize the observed 
mass-function distribution for the long-period NGC 188 blue straggler binaries when companions are drawn from the evolved 
tertiary-mass distribution.

\section{Identification of Field Blue Stragglers}

The field blue straggler sample used here is taken from the literature\cite{car05}.  Identification of blue stragglers in the Galactic field is 
not as straightforward as in open clusters because stars in the field have a wide range in age, and hence generally there is no unique main-sequence 
turnoff from which to separate normal main-sequence stars from blue stragglers.  These particular field blue stragglers were 
identified within a stellar sample limited to contain only metal-poor thick disk and halo stars, which are found to be coeval and 
of similar age to stars in globular clusters.\cite{car89}  For a given metallicity, a blue straggler is defined as being bluer than the 
main-sequence turnoff of a comparable-metallicity globular cluster.

Due to this selection technique these field blue stragglers are somewhat more metal poor and older than the blue stragglers in 
NGC 188.  Here we briefly explore the impact of these differences in the blue straggler samples.  

Metallicity is known to effect the core mass of a giant star and therefore the remnant white dwarf mass.
For the potential asymptotic-giant star donors to the long-period blue stragglers studied here, the difference in Carbon-Oxygen
white dwarf mass is at most 0.15~\Msolar.\cite{men08}  This small potential difference in companion mass does 
not effect our conclusions.  

Older main-sequence stars are less massive than their higher-mass counterparts (as high-mass stars evolve more quickly), 
and therefore, given their observational definition relative to the turnoff, older blue stragglers can be less massive as well.  
The difference in the main-sequence turnoff mass between a 7 Gyr cluster (like NGC 188) and a 
12 Gyr cluster (like a typical globular cluster) is about 0.1~\Msolar.\cite{mar08}  This small difference in blue straggler mass will 
not effect our comparison.

Finally at a given mass, lower metallicity stars also have smaller radii, due to the lower opacity level,
which in turn results in lower mass-loss rates from stellar winds.\cite{wil00}  Both of these properties may effect the 
probability of undergoing stable mass transfer in a binary system.
For instance, it has been suggested from theory that Case C mass transfer between intermediate-mass stars of lower metallicity 
may occur for a broader period range than for those of higher metallicity.\cite{dem08} 
However the complete dependence of metallicity on 
mass transfer for stars of roughly solar mass has not been fully explored.

These minor selection effects do not impact the comparison shown in this paper or the conclusions drawn from this analysis.

\end{document}